\documentclass[10pt,conference,letterpaper]{IEEEtran}

\IEEEoverridecommandlockouts

\usepackage{times,amsmath,epsfig,amssymb,graphicx,amsfonts,amsthm}
\usepackage{empheq}
\usepackage{graphicx}
\usepackage{dsfont}
\usepackage{mathrsfs}  
\usepackage{flushend}
\usepackage{psfrag}
\usepackage{tabu}
\usepackage{fge}
\usepackage{cite}
\usepackage{amsmath}
\usepackage{amsthm}
\usepackage{amsfonts}   
\usepackage{amssymb}    
\usepackage{mathrsfs}
\usepackage{caption}
\usepackage{setspace}
\usepackage{tikz}
\usepackage{dblfloatfix} 
\usepackage{caption}
\usepackage{subcaption}
\usepackage{tkz-tab}
\usepackage{color,soul}
\usepackage{algorithm}
\usepackage{algcompatible}
\usepackage{color}
\usepackage{mathrsfs}
\usepackage{mathtools}
\mathtoolsset{showonlyrefs}
\usepackage{widetext}
\usepackage{epstopdf}

\newtheorem{mydef}{Definition}

\newtheorem{mytheorem}{Theorem}

\newtheorem{myexample}{Example}

\DeclareMathAlphabet\mathbfcal{OMS}{cmsy}{b}{n}

\newcommand\T{\rule{0pt}{15pt}}

\newcounter{ale}

\newenvironment{liste}{\begin{itemize}}{\end{itemize}}
\newcommand{\aliste}{\begin{liste} \setcounter{ale}{1}}
\newcommand{\zliste}{\end{liste}}

\title{\LARGE {\bf Learning without Recall by Random Walks on Directed Graphs}\vspace*{-10pt}}

\author{M. A. Rahimian, S. Shahrampour, A. Jadbabaie{\small $~^{*}$ } 
\thanks{$^{*}$  The authors are with the Department of Electrical and Systems Engineering, University of Pennsylvania, Philadelphia, PA 19104-6228 USA (email: {\fontsize{8}{8}\selectfont\ttfamily\upshape jadbabai@seas.upenn.edu}). This work was supported by ARO MURI W911NF-12-1-0509.}
\vspace*{-10pt}}
\begin{document}

\maketitle

\begin{abstract} We consider a network of agents that aim to learn some unknown state of the world using private observations and exchange of beliefs. At each time, agents observe private signals generated based on the true unknown state. Each agent might not be able to distinguish the true state based only on her private observations. This occurs when some other states are observationally equivalent to the true state from the agent's perspective. To overcome this shortcoming, agents must communicate with each other to benefit from local observations. We propose a model where each agent selects one of her neighbors randomly at each time. Then, she refines her opinion using her private signal and the prior of that particular neighbor. The proposed rule can be thought of as a Bayesian agent who cannot recall the priors based on which other agents make inferences. This learning without recall approach preserves some aspects of the Bayesian inference while being computationally tractable. By establishing a correspondence with a random walk on the network graph, we prove that under the described protocol, agents learn the truth exponentially fast in the almost sure sense. The asymptotic rate is expressed as the sum of the relative entropies between the signal structures of every agent weighted by the stationary distribution of the random walk. 

\end{abstract}

\section{Introduction \& Background}


Distributed estimation and detection problems have been interesting subject of study in a variety of disciplines, ranging from control theory to statistics, economics, and signal processing \cite{BorkarVaraiya82,tsitsiklis1993decentralized,lopes2007incremental,bullo2009distributed,kar2009distributed,drakopoulos2013learning,shahrampour2013online,alanyali2004distributed}. In the distributed detection problem, each agent observes a sequence of independent and identically distributed (i.i.d.) private signals generated according to the true (unknown) state. Suppose that each agent forms a belief about the true state, represented by a discrete probability distribution over a {\it finite} state space, and sequentially performs the Bayes' rule to her observations at each step. It is well-known \cite{BlackwellDubins62,LehrerSmorodinsky96} that the beliefs formed in the above manner constitute a bounded martingale and converge to a limiting distribution as the number of observations tends to infinity. However, the limiting distribution is not necessarily concentrated on the truth, in which case the agent fails to learn the true state asymptotically. In fact, in many scenarios, the agent faces an identification problem where there are states (other than the true state) that are observationally equivalent to the true state. In other words, these states induce the same distribution on her sequence of privately observed signals. Therefore, rational agents  communicate in a social network to distinguish the truth by relying on local observations. This leads to the problem of social learning that is a classical focus of behavioral microeconomic theory \cite{Chamley2004,Jackson2008}, also studied in the context of distributed estimation and statistical learning theory \cite{BorkarVaraiya82,kamyar_CDC_2010}.

On the other hand, sequentially applying Bayes' rule in networks can become computationally intractable since the global network structure is not available to individuals. This origins from the fact that agents should use their local data that is increasing with time, and infer about the global signal structure. Therefore, the analysis of rational behavior in networks is an important problem in Bayesian economics, and has attracted a considerable attention \cite{mueller2013general,MosselSlyTamuz14}. On the other side of the spectrum lie the works such as \cite{Jadbabaie2012210,shahin_CDC_2013,6874893,7171178} which aim to study the problem of learning in networks via iterative applications of non-Bayesian rules. These updates provide the asymptotic properties of learning and consensus under certain conditions. More recently, some works (e.g. see \cite{shahrampour2014distributed,7172262}) have also provided the {\it non-asymptotic} analysis of the problem.

 In this paper, we study a distributed learning model where each agent observes a sequence of independent and identically distributed private signals. The structure of the network (which we assume to be strongly connected) is preset in the sense that all agents know their local neighborhood before the learning process. However, they do not necessarily contact all their neighbors every time. At every epoch of time, each agent randomly selects one neighbor, and uses her neighbor's prior (rather than herself) in the form of the Bayes' update together with her private signal at that instant of time. This can be seen as a \emph{learning without recall} rule where agents randomly pick their priors from their local neighborhood. Intuitively, asymptotic learning occurs since each agent performs a {\it random walk} over a strongly connected graph and picks up the  privately observed signals of the nodes as they are hit by the random walk. We show that the learning rate for such an agent is exponentially fast with an asymptotic rate that can be expressed as the weighted sum of the relative entropies between the likelihood structures of each agent under various states of the world, and the weights are the their probabilities in the stationary distribution of the random walk. In many distributed learning models over random and switching networks, agents must have positive self-reliant at any time. One can observe this condition, for instance, in gossip algorithms \cite{boyd2006randomized} and ergodic stationary processes \cite{tahbaz2010consensus}. An interesting and subtle point in our communication structure is the relaxation of this condition, as our agents rely entirely on the beliefs of their neighbors every time that they select a neighbor to gossip with. Moreover, unlike the majority of results that rely on the convergence properties of products of stochastic matrices and are applicable only to irreducible and aperiodic communication matrices, cf. \cite[Proporition 1]{GolubWisdomCrowd}; our results do not require the transition probability matrix to be aperiodic. This is because our proof of convergence relies on the ergodic theorem for the almost-sure convergence of the long-run fraction of time that is spent in any state of a Markov chain; and it holds true for any irreducible, positive-recurrent chain, and in particular any irreducible, finite-state chain \cite[Theorem 1.5.6]{norris1999markov}. It is further true that such a chain has a unique stationary distribution \cite[Theorem 1.7.7]{norris1999markov}, which we use to characterize the almost-sure exponentially fast asymptotic rate of convergence under our proposed distributed learning model.

The remainder of this paper is organized as follows. The modeling and formulation are set forth in Section \ref{sec:model}, where we present the signal and belief structures and their evolution. We end section \ref{sec:model} by a description of learning without recall updates in sparse structures where the neighborhood of each agent has at most one node. Next in Section \ref{sec:general} we show how the preceding updates can be used even when the agents' neighborhood are not singletons. This achieved by implementing a gossip-like procedure where a single neighbors is chosen randomly at every time-step and communications are performed with only one neighbor at a time. We study the properties of convergence and learning under this procedure and show a correspondence with random walks on directed graphs that simplifies our analysis. An illustration is provided at the end of Section~\ref{sec:general}, and the paper is concluded by Section~\ref{sec:conclusions}.

\section{The Model}\label{sec:model}

\paragraph*{Notation}  Throughout the paper, $\mathbb{R}$ is the set of real numbers, $\mathbb{N}$ denotes the set of all natural numbers, and $\mathbb{W} = \mathbb{N}\cup \{0\}$. For $n \in \mathbb{N}$ a fixed integer the set of integers $\{1,2,\ldots,n\}$ is denoted by $[n]$, while any other set is represented by a calligraphic capital letter. The cardinality of a set $\mathcal{X}$, which is the number of its elements, is denoted by $\mid\mathcal{X}\mid$, and $\mathscr{P}(\mathcal{X}) = \{ \mathcal{M}; \mathcal{M} \subset \mathcal{X} \}$ denotes the power-set of $\mathcal{X}$, which is the set of all its subsets. The difference of two sets $\mathcal{X}$ and $\mathcal{Y}$ is defined by $\mathcal{X} \fgebackslash \mathcal{Y}:=\left\{x;x \in \mathcal{X} \mbox{ and } x \notin \mathcal{Y}\right\}$. Boldface letters denote random variables. 

We consider a network of $n$ agents that interact according to a directed graph $\mathcal{G} = ([n],\mathcal{E})$, where $\mathcal{E} \subset [n] \times [n]$ is the set of directed edges. Each agent is labeled with an element of the set $[n]$. $\mathcal{N}(i) = \{j \in [n]; (j,i) \in \mathcal{E}\}$ is  the neighborhood of agent $i$ which is the set of all agents whose beliefs can be observed by agent $i$. We let $\deg(i) = \mid\mathcal{N}(i)\mid$ be the degree of node $i$ corresponding to the number of agent $i$'s neighbors. 

\paragraph*{The Environment} We denote by $\Theta$ the set of states of the world which has a finite cardinality. Also, $\Delta\Theta$ represents the space of all probability measures on the set $\Theta$. Each agent's goal is to decide amongst the finitely many possibilities in the state space $\Theta$. A random variable $\boldsymbol{\theta}$ is chosen randomly from $\Theta$ by the nature and according to the probability measure $\nu(\mathord{\cdot}) \in \Delta\Theta$, which satisfies $\nu(\hat{\theta}) > 0,\forall \hat{\theta} \in \Theta$ and is referred to as the common prior. For each agent $i$, there exists a finite signal space denoted by $\mathcal{S}_i$, and given $\boldsymbol{\theta}$, $\ell_i(\mathord{\cdot}\mid\boldsymbol{\theta})$ is a probability measure on $\mathcal{S}_i$, which is referred to as the \emph{signal structure} or \emph{likelihood function} of agent $i$. Furthermore, $(\Omega,\mathscr{F},\mathbb{P})$ is a probability triplet, where 
\begin{align}
\Omega = \Theta \times {\left(\prod_{i\in[n]}\mathcal{S}_i\right)}^{\mathbb{W}}, 
\end{align}
is an infinite product space with a general element $\omega = (\theta;(s_{1,0},\ldots,s_{n,0}),(s_{1,1},\ldots,s_{n,1}),\ldots)$ and the associated sigma field $\mathscr{F} = \mathscr{P}(\Omega)$.  $\mathbb{P}(\mathord{\cdot})$ is the probability measure on $\Omega$  which assigns probabilities consistently with the common prior $\nu(\mathord{\cdot})$ and the likelihood functions $\ell_i(\mathord{\cdot}\mid\boldsymbol{\theta}), i \in [n]$. Conditioned on $\boldsymbol{\theta}$, the random vectors $\{(\mathbf{s}_{1,t},\ldots,\mathbf{s}_{n,t}),t\in\mathbb{W}\}$ are independent. $\mathbb{E}\{\mathord{\cdot}\}$ is the expectation operator, which represents integration with respect to $d\mathbb{P}(\omega)$. 

\paragraph*{Signals} Let $t \in \mathbb{W}$ denote the time index and for each agent $i$, define $\{\mathbf{s}_{i,t},t\in \mathbb{W}\}$ to be a sequence of independent and identically distributed random variables with the probability mass function $\ell_i(\mathord{\cdot}\mid\boldsymbol{\theta})$; this sequence represents the private observations made by agent $i$ at each time period $t$. The privately observed signals are independent and identically distributed over time, but they could be correlated across the agents.

\paragraph*{Beliefs} We let ${\boldsymbol\mu}_{i,t}(\mathord{\cdot})$ represent the \emph{opinion} or \emph{belief} at time $t$ of agent $i$ about the realized value of $\boldsymbol{\theta}$. In other words, ${\boldsymbol\mu}_{i,t}(\mathord{\cdot})$ is a probability distribution on the set $\Theta$ at any time $t$ formed by agent $i$. Note the randomness of ${\boldsymbol\mu}_{i,t}(\mathord{\cdot})$ due to its dependence on the random observations of the agent. The goal is to study asymptotic learning, i.e. for each agent to learn the true realized value $\theta \in \Theta$ of $\boldsymbol{\theta}$ asymptotically. This amounts to having $\boldsymbol\mu_{i,t}(\mathord{\cdot})$ converge to a point mass centered at $\theta$, where the convergence could be in probability or in the stronger almost sure sense that we use in this work.

At $t= 0$ the value $\boldsymbol{\theta} = \theta$ is selected by nature. Followed by that, $\mathbf{s}_{i,0}$ for each $i\in [n]$ is realized and observed by agent $i$. Then the agent forms an initial Bayesian opinion ${\boldsymbol\mu}_{i,0}(\mathord{\cdot})$ about the value of $\theta$. Given $\mathbf{s}_{i,0}$, and using the Bayes' rule for each agent $i\in[n]$, the initial belief in terms of the observed signal $\mathbf{s}_{i,0}$ is given by:
\begin{equation}
{\boldsymbol\mu}_{i,0}(\hat{\theta}) = \frac{ \nu(\hat{\theta})\ell_i(\mathbf{s}_{i,0}\mid \hat{\theta} )} {\displaystyle\sum_{\tilde{\theta} \in \Theta} \nu(\tilde{\theta})\ell_i(\mathbf{s}_{i,0} \mid \tilde{\theta} )}.
\label{eq:bayes1}
\end{equation} 
Afterwards, at any time $t$ each agent $i$ observes the realized value of $\mathbf{s}_{i,t}$ as well as the current belief of one of her neighbors ${\boldsymbol\mu}_{k,t-1}(\mathord{\cdot})$, where $k$ is selected randomly from $ \mathcal{N}(i)$.  She then forms a refined opinion ${\boldsymbol\mu}_{i,t}(\mathord{\cdot})$ by incorporating all the data that have been made available to her by the time $t$. We elaborate on the update rule in the following.

\section{Combined Gossip and without Recall Updates: Signals Picked up in a Random Walk }\label{sec:general}

Consider a digraph $\mathcal{G}$ satisfying $\deg(i)\in\{0,1\},\forall i \in [n]$. For this class of networks which include directed circles and rooted trees in \cite{LWCcircleTree,rahimian2014non}, the authors propose to use the Bayesian update 
\begin{align}
{\boldsymbol\mu}_{i,t}(\hat{\theta}) = \frac{ {\boldsymbol\mu}_{i,t-1}(\hat{\theta})\ell_i(\mathbf{s}_{i,t}\mid\hat{\theta})} {\displaystyle\sum_{\tilde{\theta} \in \Theta}{\boldsymbol\mu}_{i,t-1}(\tilde{\theta})\ell_i(\mathbf{s}_{i,t}\mid\tilde{\theta})}, \forall \hat{\theta} \in \Theta,
\label{eq:bayesSINGLE}
\end{align} if $\deg(i) = 0 $; and else to use 
\begin{align}
{\boldsymbol\mu}_{i,t}(\hat{\theta}) = \frac{ {\boldsymbol\mu}_{j,t-1}(\hat{\theta})\ell_i(\mathbf{s}_{i,t}\mid\hat{\theta})} {\displaystyle\sum_{\tilde{\theta} \in \Theta}{\boldsymbol\mu}_{j,t-1}(\tilde{\theta})\ell_i(\mathbf{s}_{i,t}\mid\tilde{\theta})}, \forall \hat{\theta} \in \Theta,
\label{eq:bayesSINGLE_neighborReplaced}
\end{align} where $j\in [n]$ is the unique vertex $j \in \mathcal{N}(i)$. These updates are a special case of the \emph{Learning without Recall} rules that are developed in a companion paper, and they can describe the behavior of Rational but Memoryless agents who share a common prior $\nu(\mathord{\cdot})$ and always interpret their current and observed beliefs as having stemmed from this common prior, thus ignoring their entire history of past observations.

Here we propose the application of the Learning without Recall updates that we described in the previous section to general networks, by requiring that at every time step $t$, node $i$ make a random choice from her set of neighbors $\mathcal{N}(i)$ and uses that choice for the unique $j$ in  \eqref{eq:bayesSINGLE_neighborReplaced}. To this end, let $\boldsymbol\sigma_t\in\Pi_{i\in[n]}\mathcal{N}(i), t \in \mathbb{N}$ be a sequence of independent and identically distributed random vectors such that $\forall t \in \mathbb{N}$, $\boldsymbol\sigma_{t,i} \in \mathcal{N}(i)$ is that neighbor of $i$ which she chooses to communicate with at time $t$. Hence, for all $t$ and any $i$, \eqref{eq:bayesSINGLE_neighborReplaced} becomes
\begin{align}
{\boldsymbol\mu}_{i,t}(\hat{\theta}) = \frac{ {\boldsymbol\mu}_{_{\boldsymbol\sigma_{t,i}},t-1}(\hat{\theta})\ell_i(\mathbf{s}_{i,t}\mid\hat{\theta})} {\displaystyle\sum_{\tilde{\theta} \in \Theta}{\boldsymbol\mu}_{_{\boldsymbol\sigma_{t,i}},t-1}(\tilde{\theta})\ell_i(\mathbf{s}_{i,t}\mid\tilde{\theta})}, \forall \hat{\theta} \in \Theta.
\label{eq:bayesSINGLE_neighborReplacedGossip}
\end{align} To proceed, annex the random choice of neighbors for every node $i \in [n]$ and all times $t \in \mathbb{N}$ to the original probability space $(\Omega,\mathscr{F},\mathbb{P})$ specified in Section~\ref{sec:model}; and for $t\in \mathbb{N}$ arbitrary, let $\mathbb{P}\{{\boldsymbol\sigma_{t,i}}=j\} = p_{i,j}>0$. Wherefore, $\sum_{j\in\mathcal{N}(i)}p_{i,j} = 1-p_{i,i}\leq1$, and $p_{i,j} = 0$ whenever $j \not\in \mathcal{N}(i)\cup\{i\}$. Let $P$ be the row stochastic matrix whose $(i,j)$-th entry is equal to $p_{i,j}$. Let $\mathds{1}_{\{{\boldsymbol\sigma_{t,i}} = j\}} =1$ if ${\boldsymbol\sigma_{t,i}} = j$ and $\mathds{1}_{\{{\boldsymbol\sigma_{t,i}} = j\}} =0$ otherwise. Then \eqref{eq:bayesSINGLE_neighborReplacedGossip} can be written as 
\begin{align}
{\boldsymbol\mu}_{i,t}(\hat{\theta}) &= \sum^{n}_{j=1}\mathds{1}_{\{{\boldsymbol\sigma_{t,i}} = j\}}\frac{ {\boldsymbol\mu}_{j,t-1}(\hat{\theta})\ell_i(\mathbf{s}_{i,t}\mid\hat{\theta})} {\displaystyle\sum_{\tilde{\theta} \in \Theta}{\boldsymbol\mu}_{j,t-1}(\tilde{\theta})\ell_i(\mathbf{s}_{i,t}\mid\tilde{\theta})} \\
& = \ell_i(\mathbf{s}_{i,t}\mid\hat{\theta}) \prod^{n}_{j=1} {\left(\frac{ {\boldsymbol\mu}_{j,t-1}(\hat{\theta})} {\displaystyle\sum_{\tilde{\theta} \in \Theta}{\boldsymbol\mu}_{j,t-1}(\tilde{\theta}) \ell_i(\mathbf{s}_{i,t} \mid \tilde{\theta})} \right)}^{\mathds{1}_{\{{\boldsymbol\sigma_{t,i}} = j\}}}  \label{eq:bayesSINGLE_neighborReplacedGossipIndicator}
\end{align}

To analyze the propagation of beliefs under \eqref{eq:bayesSINGLE_neighborReplacedGossip} we form the belief ratio
\begin{align}
\frac{{\boldsymbol\mu}_{i,t}(\check{\theta})}{{\boldsymbol\mu}_{i,t}({\theta})}  &= \frac{\ell_i(\mathbf{s}_{i,t}\mid\check{\theta})}{\ell_i(\mathbf{s}_{i,t}\mid {\theta})} \prod^{n}_{j=1}{\left(\frac{ {\boldsymbol\mu}_{j,t-1}(\check{\theta})} { {\boldsymbol\mu}_{j,t-1}({\theta})}\right)}^{\mathds{1}_{\{{\boldsymbol\sigma_{t,i}} = j\}}}
\label{eq:ratioEvolution}
\end{align} for any false state $\check{\theta}$ $\in$ $\Theta\fgebackslash \{\theta\}$ and each agent $i\in[n]$ at all times $t \in \mathbb{N}$. The above has the advantage of removing the normalization factor in the dominator out of the picture; thence, focusing instead on the evolution of belief ratios. To proceed, we take the logarithms of both sides in \eqref{eq:ratioEvolution} to obtain 
\begin{align}
\log\left(\frac{{\boldsymbol\mu}_{i,t}(\check{\theta})}{{\boldsymbol\mu}_{i,t}({\theta})}\right) & =  \log\left(\frac{\ell_i(\mathbf{s}_{i,t}\mid\check{\theta})}{\ell_i(\mathbf{s}_{i,t}\mid {\theta})}\right) \label{eq:ratioEvolutionLOG} \\ & + \sum^{n}_{j=1}{\mathds{1}_{\{{\boldsymbol\sigma_{t,i}} = j\}}}\log {\left(\frac{ {\boldsymbol\mu}_{j,t-1}(\check{\theta})} { {\boldsymbol\mu}_{j,t-1}({\theta})}\right)}
\end{align}


Next we can iterate \eqref{eq:ratioEvolutionLOG} to replace for $({\boldsymbol\mu}_{j,t-1}(\check{\theta})/ {\boldsymbol\mu}_{j,t-1}({\theta}))$ and so on, from which we get \eqref{eq:signalRatioSumFormula}  at the top of next page. Also note,

\begin{align}
\sum^{n}_{i_1=1}\ldots\sum^{n}_{i_t=1}\mathds{1}_{\{{\boldsymbol\sigma_{t,i}} = i_1\}} \ldots \mathds{1}_{\{\boldsymbol\sigma_{1,i_{t-1}} = i_t\}} =  1,
\end{align} almost surely, and in fact every where on $\Omega$, so that the initial prior belief ratio $\log (\nu(\check{\theta}) / \nu(\theta) )$ always appears in the summation \eqref{eq:signalRatioSumFormula}, and it simplifies as in \eqref{eq:signalRatioSumFormula2} at the top of next page.

\begin{figure*}
\begin{align}
\log\left(\frac{{\boldsymbol\mu}_{i,t}(\check{\theta})}{{\boldsymbol\mu}_{i,t}({\theta})}\right)  &= \log\left(\frac{\ell_i(\mathbf{s}_{i,t}\mid\check{\theta})}{\ell_i(\mathbf{s}_{i,t}\mid {\theta})}\right) + \sum^{n}_{i_1=1} \mathds{1}_{\{{\boldsymbol\sigma_{t,i}} = i_1\}}  \log\left(\frac{\ell_{i_1}(\mathbf{s}_{i_1,t-1}\mid\check{\theta})}{\ell_{i_1}(\mathbf{s}_{i_1,t-1}\mid {\theta})}\right) \\ & +  \mathds{1}_{\{{\boldsymbol\sigma_{t,i}} = i_1\}} \sum^{n}_{i_2=1} \mathds{1}_{\{{\boldsymbol\sigma_{t-1,i_1}} = i_2\}} \log {\left(\frac{ {\ell}_{i_2}(\mathbf{s}_{i_2,t-2} \mid \check{\theta})} { {\ell}_{i_2}(\mathbf{s}_{i_2,t-2} \mid {\theta})}\right)}\\ 
& +  \mathds{1}_{\{{\boldsymbol\sigma_{t,i}} = i_1\}} \mathds{1}_{\{{\boldsymbol\sigma_{t-1,i_1}} = i_2\}} \sum^{n}_{i_3=1} \mathds{1}_{\{\boldsymbol\sigma_{t-2,i_2} = i_3\}}  \log {\left(\frac{ {\ell}_{i_3}(\mathbf{s}_{i_3,t-3} \mid \check{\theta})} { {\ell}_{i_3}(\mathbf{s}_{i_3,t-3} \mid {\theta})}\right)} + \ldots \\ &+  \mathds{1}_{\{{\boldsymbol\sigma_{t,i}} = i_1\}} \mathds{1}_{\{{\boldsymbol\sigma_{t-1,i_1}} = i_2\}}...\mathds{1}_{\{\boldsymbol\sigma_{1,i_{t-2}} = i_{t-1}\}}\sum^{n}_{i_t=1} \mathds{1}_{\{\boldsymbol\sigma_{1,i_{t-1}} = i_t\}} \left\{ \log {\left(\frac{ {\ell}_{i_t}(\mathbf{s}_{i_t,0} \mid \check{\theta})} { {\ell}_{i_t}(\mathbf{s}_{i_t,0} \mid {\theta})}\right)} +\log {\left(\frac{ \nu(\check{\theta})} { \nu(\theta)}\right)}   \right\}  \label{eq:signalRatioSumFormula}
\end{align}

\hrulefill

\begin{align}
& \log\left(\frac{{\boldsymbol\mu}_{i,t}(\check{\theta})}{{\boldsymbol\mu}_{i,t}({\theta})}\right)   = \log\left(\frac{\ell_i(\mathbf{s}_{i,t}\mid\check{\theta})}{\ell_i(\mathbf{s}_{i,t}\mid {\theta})}\right) + \log {\left(\frac{ \nu(\check{\theta})} { \nu(\theta)}\right)} + \sum^{n}_{i_1=1}{\mathds{1}_{\{{\boldsymbol\sigma_{t,i}} = i_1\}}} \{ \log {\left(\frac{\ell_{i_1}(\mathbf{s}_{i_1,t-1}\mid\check{\theta})}{\ell_{i_1}(\mathbf{s}_{i_1,t-1}\mid {\theta})}\right)}  \\ & + \sum^{n}_{i_2=1}{\mathds{1}_{\{{\boldsymbol\sigma_{t,i_1}} = i_2\}}} \{ \log {\left(\frac{\ell_{i_2}(\mathbf{s}_{i_2,t-2}\mid\check{\theta})}{\ell_{i_2}(\mathbf{s}_{i_2,t-2}\mid {\theta})}\right)}  + \ldots  + \sum^{n}_{i_\tau=1} {\mathds{1}_{\{{\boldsymbol\sigma_{t-\tau+1,i_{\tau-1}}} = i_\tau\}}} \{ \log {\left(\frac{\ell_{i_\tau}(\mathbf{s}_{i_\tau,t-\tau}\mid\check{\theta})}{\ell_{i_\tau}(\mathbf{s}_{i_\tau,t-\tau}\mid {\theta})}\right)} \\ & + \ldots  + \sum_{i_{t-1} = 1}^{n} \mathds{1}_{\{\boldsymbol\sigma_{1,i_{t-2}} = i_{t-1}\}} \{ \log {\left(\frac{ {\ell}_{i_{t-1}}(\mathbf{s}_{i_{t-1},1} \mid \check{\theta})} { {\ell}_{i_{t-1}}(\mathbf{s}_{i_{t-1},1} \mid {\theta})}\right)} + \sum^{n}_{i_t=1} \mathds{1}_{\{\boldsymbol\sigma_{1,i_{t-1}} = i_t\}}  \log {\left(\frac{ {\ell}_{i_t}(\mathbf{s}_{i_t,0} \mid \check{\theta})} { {\ell}_{i_t}(\mathbf{s}_{i_t,0} \mid {\theta})}\right)}\} \ldots \} \label{eq:signalRatioSumFormula2}
\end{align}

\hrulefill

\end{figure*}

We now claim that whenever $t \to \infty$ and the network graph $\mathcal{G}$ is strongly connected, with $\mathbb{P}$-probability one the likelihood ratios of private signals from any node $m\in[n]$ appears in the summation \eqref{eq:signalRatioSumFormula2} as $\ell_{m}(\mathbf{s}_{m,t-\tau}\mid \check{\theta})/\ell_{m}(\mathbf{s}_{m,t-\tau}\mid {\theta})$ for infinitely many values of $\tau$. The gist of the proof is in realizing the correspondence between the summation \eqref{eq:signalRatioSumFormula} and a random walk on the directed graph $\mathcal{G}$ that starts at time $t$ on node $i$, proceeds in the reversed time direction, and terminates at time zero. The jumps in this random walk are made from each node $i$ to one of her in-neighbors $j\in\mathcal{N}(i)$ and in accordance with the probabilities $p_{i,j}$ specified by matrix $P = [p_{i,j}]$. Indeed, we can denote the random sequence of nodes that are hit by this random walk as $({i},\mathbf{i}_{1} , \ldots , \mathbf{i}_t)$ where the random variables $\mathbf{i}_{\tau}\in[n]$, $\tau \in [t]$ are defined recursively by $ \mathbf{i}_1:=\boldsymbol\sigma_{t,i}$, $ \mathbf{i}_2:=\boldsymbol\sigma_{t-1,\boldsymbol\sigma_{t,\mathbf{i}_1}}$, $ \mathbf{i}_3:=\boldsymbol\sigma_{t-2,\boldsymbol\sigma_{t-1,\mathbf{i}_2}}$, $\ldots$, $\mathbf{i}_t:=\boldsymbol\sigma_{1,\boldsymbol\sigma_{2,\mathbf{i}_{t-1}}}$ . Whence \eqref{eq:signalRatioSumFormula2} is written succinctly as
\begin{align}
 \log\left(\frac{{\boldsymbol\mu}_{i,t}(\check{\theta})}{{\boldsymbol\mu}_{i,t}({\theta})}\right)   &= \log\left(\frac{\ell_i(\mathbf{s}_{i,t}\mid\check{\theta})}{\ell_i(\mathbf{s}_{i,t}\mid {\theta})}\right) +  \log {\left(\frac{ \nu(\check{\theta})} { \nu(\theta)}\right)} \\ &+ \sum^{t}_{\tau=1} \log {\left(\frac{\ell_{\mathbf{i}_\tau}(\mathbf{s}_{\mathbf{i}_\tau ,t-\tau}\mid\check{\theta})}{\ell_{\mathbf{i}_\tau}(\mathbf{s}_{\mathbf{i}_\tau,t-\tau}\mid {\theta})}\right)}.
\label{eq:signalRatioSumFormula3}
\end{align} As $t\to\infty$, the sequence $\mathbf{i}_{\tau},\tau\in\mathbb{N}$ forms a Markov process with transition matrix $P$. Given \eqref{eq:signalRatioSumFormula3}, our claim can be restated as that for every $m\in[n]$ and as $t\to\infty$ there are infinitely many values of $\tau\in\mathbb{N}$ for which $\mathbf{i}_\tau = m$, and it is true because in a finite state Markov chain with transition matrix $P$ every state is persistent (recurrent) and will be hit infinitely many times provided that the directed graph $\mathcal{G}$ is strongly connected \cite[Theorem 1.5.6]{norris1999markov}, i.e. we have that $\forall m \in [n]$,
\begin{align}
\mathbb{P}\{\mathbf{i}_\tau = m, \mbox{for infinitely many $\tau$}\} = 1. 
\end{align} For any agent $m\in[n]$ let $\mathbfcal{T}_m := \{\boldsymbol\tau_{m,j}, j \in\mathbb{N}\}$ be the sequence of stopping times that record the first, second and so on passage times of node $m$ by the process $\mathbf{i}_{\tau},\tau\in\mathbb{N}$. That is we have $\boldsymbol\tau_{m,1} = \inf\{\tau\in\mathbb{N}:\mathbf{i}_{\tau} = m\}$ and for $j>1$, $\boldsymbol\tau_{m,j} = \inf\{\tau> \boldsymbol\tau_{m,j-1}:\mathbf{i}_{\tau} = m\}$. Using the above notation, \eqref{eq:signalRatioSumFormula3} can be rewritten as 
\begin{align}
 \log\left(\frac{{\boldsymbol\mu}_{i,t}(\check{\theta})}{{\boldsymbol\mu}_{i,t}({\theta})}\right) &= \log\left(\frac{\ell_i(\mathbf{s}_{i,t}\mid\check{\theta})}{\ell_i(\mathbf{s}_{i,t}\mid {\theta})}\right) +  \log {\left(\frac{ \nu(\check{\theta})} { \nu(\theta)}\right)} \\ &+ \sum_{m=1}^{n}\sum_{\substack{\tau\in\mathbfcal{T}_m ,\\ \tau\leq t}} \log {\left(\frac{\ell_{m}(\mathbf{s}_{m,t-\tau}\mid\check{\theta})}{\ell_{m}(\mathbf{s}_{m,t-\tau}\mid {\theta})}\right)}.
\label{eq:signalRatioSumFormula4}
\end{align} On the other hand, note that $\log {\left( {\ell_{m}(\mathbf{s}_{m,t-\boldsymbol\tau_m,j}\mid\check{\theta})}/{\ell_{m}(\mathbf{s}_{m,t-\boldsymbol\tau_m,j}\mid {\theta})}\right)}$, $j\in \mathbb{N}$ is a sequence of independent and identically distributed signals, so that by the strong of large numbers we obtain that with $\mathbb{P}$-probability one,
\begin{align}
&\lim_{n\to\infty}\frac{1}{n}\sum_{j=1}^{n}\log {\left(\frac{\ell_{m}(\mathbf{s}_{m,t-\boldsymbol\tau_m,j}\mid\check{\theta})}{\ell_{m}(\mathbf{s}_{m,t-\boldsymbol\tau_m,j}\mid {\theta})}\right)} \\ &= \mathbb{E}\log {\left(\frac{\ell_{m}(\mathbf{s}_{m,0}\mid\check{\theta})}{\ell_{m}(\mathbf{s}_{m,0}\mid {\theta})}\right)}  :=  - D_{KL}\left(\ell_m( \mathord{\cdot} |\theta) \| \ell_m( \mathord{\cdot} |\check{\theta}) \right) \leqslant 0,
\label{eq:SLLN}
\end{align} where the non-positivity follows from the information inequality for the Kullback-Leibler divergence $D_{KL}\left(\mathord{\cdot}|| \mathord{\cdot} \right)$ and is strict whenever $\ell_m( \mathord{\cdot} |\check{\theta}) \not\equiv \ell_m( \mathord{\cdot} |\theta)$, i.e. $\exists s \in \mathcal{S}_i$ such that $\ell_i( s |\check{\theta}) \neq \ell_i( s |\theta)$ \cite[Theorem 2.6.3]{cover2006elements}. Note that whenever $\ell_i( \mathord{\cdot} |\hat{\theta}) \equiv \ell_i( \mathord{\cdot} |\theta)$ or equivalently $D_{KL}\left(\ell_i( \mathord{\cdot} |\hat{\theta})\| \ell_i( \mathord{\cdot} |\theta)) \right) = 0$, then the two states $\hat{\theta}$ and $\theta$ are statically indistinguishable to agent $i$. In other words, there is no way for agent $i$ to differentiate $\hat{\theta}$ from $\theta$ based only on her private signals. This follows from the fact that both $\theta$ and $\hat{\theta}$ induce the same probability distribution on her sequence of observed i.i.d. signals. On the other hand, having $D_{KL}\left(\ell_m( \mathord{\cdot} |\theta) \| \ell_m( \mathord{\cdot} |\check{\theta}) \right)<0$ for some agent $m\in[n]$ would ensure per \eqref{eq:SLLN} and persistence of state $m$ that with $\mathbb{P}$-probability one,
\begin{align}
\sum_{\substack{\tau\in\mathbfcal{T}_m ,\\ \tau\leq t}} \log {\left(\frac{\ell_{m}(\mathbf{s}_{m,t-\tau}\mid\check{\theta})}{\ell_{m}(\mathbf{s}_{m,t-\tau}\mid {\theta})}\right)} \to -\infty
\end{align} as $t\to\infty$ in \eqref{eq:signalRatioSumFormula4}; consequently, $ \log\left({{\boldsymbol\mu}_{i,t}(\check{\theta})}/{{\boldsymbol\mu}_{i,t}({\theta})}\right) \to -\infty$ for all agent $i\in[n]$ and any such $\check{\theta}\in\Theta$, $\check{\theta}\neq\theta$. Indeed, having $ \log\left({{\boldsymbol\mu}_{i,t}(\check{\theta})}/{{\boldsymbol\mu}_{i,t}({\theta})}\right) \to -\infty$ for all $\check{\theta}\neq \theta$ is necessary and sufficient for learning, and we therefore, have the following characterization.

\begin{mydef}[Global Identifiability]\label{def:learnability} In a strongly connected topology, the true state $\theta$ is globally identifiable, if for all $\check{\theta}\neq{\theta}$ there exists some agent $m\in[n]$ such that $D_{KL}\left(\ell_m( \mathord{\cdot} |\theta) \| \ell_m( \mathord{\cdot} |\check{\theta}) \right)<0$, i.e. $m$ can distinguish between $\check{\theta}$ and ${\theta}$ based only on her private signals.
\end{mydef}

We have thus established the conditions for learning under the without recall updates in \eqref{eq:bayesSINGLE} and \eqref{eq:bayesSINGLE_neighborReplaced}, where the neighbor $j$ is chosen randomly with strictly positive probabilities specified in transition matrix $P$. We dub this procedure ``\emph{gossips without recall}'' and summarize our findings as follows:
 
\begin{mytheorem}[Almost-Sure Learning]\label{theo:learning} Under the \emph{gossips without recall} updates in a strongly connected network where the truth is globally identifiable, all agents learn the truth asymptotically almost surely.
\end{mytheorem}

We can extend the above analysis to derive an asymptotic rate of learning for the agents that is exponentially fast and is expressed as $\sum_{m=1}^{m} {\pi}_m D_{KL}\left(\ell_m( \mathord{\cdot} |\theta) \| \ell_m( \mathord{\cdot} |\check{\theta}) \right)<0$, where $\overline{\pi}:=(\pi_1,\ldots,\pi_n)$ is the stationary distribution of the transition matrix $P$, which for a strongly connected $\mathcal{G}$ is the unique probbaility distribution on $[n]$ satisfying $\overline{\pi}P = \overline{\pi}$. To see how, for each agent $m\in[n]$ and all time $t$, define $\mathbfcal{T}_m(t) := \{\boldsymbol\tau_{m,j}, j \in\mathbb{N}:\boldsymbol\tau_{m,j}\leq t\}$ and divide both sides of \eqref{eq:signalRatioSumFormula4} by $t$ to obtain
\begin{align}
\frac{1}{t} \log\left(\frac{{\boldsymbol\mu}_{i,t}(\check{\theta})}{{\boldsymbol\mu}_{i,t}({\theta})}\right) &= \frac{1}{t}\log\left(\frac{\ell_i(\mathbf{s}_{i,t}\mid\check{\theta})}{\ell_i(\mathbf{s}_{i,t}\mid {\theta})}\right) + \frac{1}{t} \log {\left(\frac{ \nu(\check{\theta})} { \nu(\theta)}\right)} \\ &+ \frac{1}{t}\sum_{m=1}^{n}\sum_{\tau\in\mathbfcal{T}_m(t)} \log {\left(\frac{\ell_{m}(\mathbf{s}_{m,t-\tau}\mid\check{\theta})}{\ell_{m}(\mathbf{s}_{m,t-\tau}\mid {\theta})}\right)}.
\label{eq:signalRatioSumFormula5}
\end{align} Upon invoking \eqref{eq:SLLN} we obtain
\begin{align}
&\lim_{t\to\infty}\frac{1}{t} \log\left(\frac{{\boldsymbol\mu}_{i,t}(\check{\theta})}{{\boldsymbol\mu}_{i,t}({\theta})}\right) = \\ &- \sum_{m=1}^{n} \lim_{t\to\infty} \frac{|\mathbfcal{T}_m(t)|}{t} D_{KL}\left(\ell_m( \mathord{\cdot} |\theta) \| \ell_m( \mathord{\cdot} |\check{\theta}) \right).
\label{eq:rate1}
\end{align} Finally the ergodic theorem ensures that the average time spent in any state $m\in[n]$ converges almost surely to its stationary probability $\pi_m$, i.e. with probability one $\lim_{t\to\infty} {|\mathbfcal{T}_m(t)|}/{t} = \pi_m$, \cite[Theorem 1.10.2]{norris1999markov}. Hence, \eqref{eq:rate1} becomes
\begin{align}
\lim_{t\to\infty}\frac{1}{t} \log\left(\frac{{\boldsymbol\mu}_{i,t}(\check{\theta})}{{\boldsymbol\mu}_{i,t}({\theta})}\right) = - \sum_{m=1}^{n} \pi_m D_{KL}\left(\ell_m( \mathord{\cdot} |\theta) \| \ell_m( \mathord{\cdot} |\check{\theta}) \right),
\label{eq:rate2}
\end{align} completing the proof for the claimed asymptotically exponentially fast rate.



\begin{figure}[H]
\centering
\begin{tikzpicture}
\tikzstyle{every node}=[draw,shape=circle];
\node (v1) at ( 0:1) {$1$};
\node (v2) at ( 72:1) {$2$};
\node (v5) at (2*72:1) {$5$};
\node (v4) at (3*72:1) {$4$};
\node (v3) at (4*72:1) {$3$};
\node (v6) at (4*72:2) {$6$};
\node (v7) at (4*72+36:1.6) {$7$};
\node (v8) at (4*72+22.39:2.49721) {$8$};

\foreach \from/\to in {v1/v2, v2/v5, v2/v3,  v3/v4, v3/v1, v3/v6, v4/v2, v1/v7, v5/v4, v7/v8}
\draw [->] (\from) -- (\to);

\draw
(v1) -- (v2)
(v2) -- (v5)
(v2) -- (v3)
(v3) -- (v4)
(v3) -- (v1)
(v3) -- (v6)
(v4) -- (v2)
(v1) -- (v7)
(v5) -- (v4)
(v7) -- (v8);
\end{tikzpicture}
\caption{Network Structure for Example 1}
\label{fig:hybrid}
\end{figure}
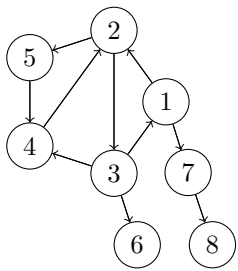

\begin{myexample}Eight Agents with Binary Signals in a Tri-State World.\end{myexample}
As an illustration consider the network of agents in Fig.~\ref{fig:hybrid} with the true state of the world being $1$, the first of the tree possible states $\Theta = \{1,2,3\}$. The likelihood structure for the first three agents is given in the table and note that none of them can learn the truth on their own; indeed, agent $3$ does not receive any informative signals and her beliefs shall never depart from their initial priors following \eqref{eq:bayesSINGLE}. We further set $l_j(\mathord{\cdot} \mid \mathord{\cdot}) \equiv l_3(\mathord{\cdot} \mid \mathord{\cdot})$ for all $j \in [8]\fgebackslash[3]$, so that all the remaining agents are also unable to infer anything about the true state of the world from their own private signals.

\begin{center}
{\tabulinesep=1.2mm 
\begin{tabu} {|c|c|c|c|} \hline 
likelihoods     & $\hat{\theta} = 1$ & $\hat{\theta} = 2$ & $\hat{\theta} = 3$ \\   \hline 
$l_1(\mathbf{s}_{1,t} = 0 \mid \hat{\theta})  $    & $\frac{1}{3}$        & $\frac{1}{3}$        & $\frac{1}{5}$    \\ \hline
$l_2(\mathbf{s}_{2,t} = 0 \mid \hat{\theta})  $    & $\frac{1}{2}$        & $\frac{2}{3}$        & $\frac{1}{2}$    \\ \hline
$l_3(\mathbf{s}_{3,t} = 0 \mid \hat{\theta})  $    & $\frac{1}{4}$        & $\frac{1}{4}$        & $\frac{1}{4}$    \\ \hline
\end{tabu}} 
\end{center} 

\begin{figure}[t!]
\centering
\includegraphics[width=87mm]{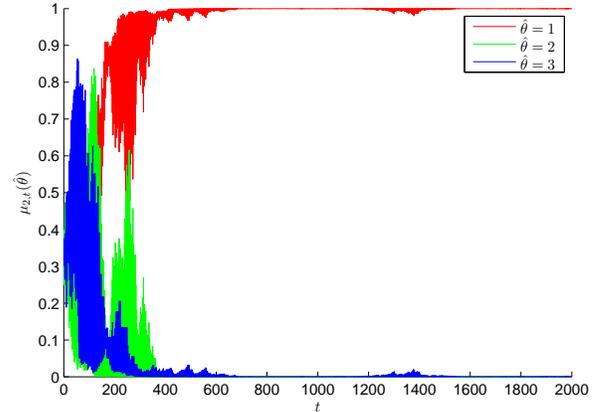}
\caption{Evolution of the second agent’s beliefs over time}
\label{shekl1}
\end{figure}

\begin{figure}[t!]
\centering
\includegraphics[width=87mm]{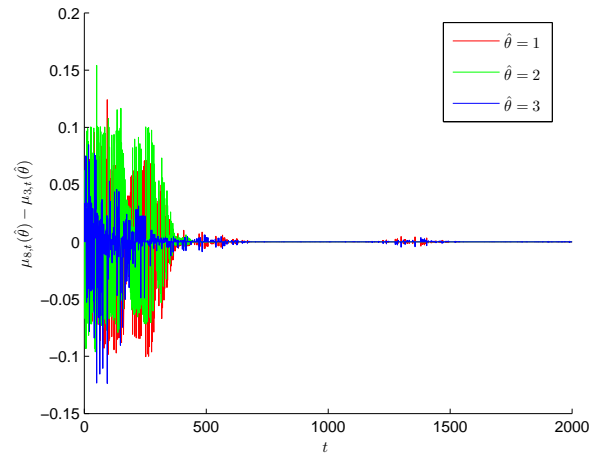}
\caption{The difference between the third and eighth agents’
beliefs over time}
\label{shekl2}
\end{figure}

Starting from a uniform common prior and following the proposed gossip without recall scheme with neighbors chosen uniformly at random, all agents asymptotically learn the true state, even though none of them can learn the true state on their own. The plots in Figs. \ref{shekl1} and  \ref{shekl2} depict the belief evolution for the second agent, as well as the difference between the beliefs for the third and eighth agents. It is further observable that all agents learn the true state at the same exponentially fast asymptotic rate of learning.

\section{Concluding Remarks}\label{sec:conclusions}

This work addressed a social and observational learning model in multi-agent networks. Agents attempt to learn some unknown state of the world which belongs to a finite state space. Conditioned on the true state, a sequence of i.i.d. private signals are generated and observed by each agent of the network. The private signals do not provide each agent with adequate information to identify the truth. Hence, agents contact their neighbors to augment their imperfect observations with those of their neighbors. In our model, every time, each agent picks a neighbor randomly and updates her belief using the prior of that particular neighbor but using the likelihood for her own private signal. The communication protocol is an instance of a \emph{learning without recall} and is implemented in such a way that signals likelihoods that comprise an agent's belief are picked up by a random walk on the network graph. We proved that agents learn the truth exponentially fast and in the almost sure sense, provided that the network is strongly and the truth is globally identifiable. The asymptotic rate is expressed as a weighted sum of the relative entropies between the signal structures of each agent, where the weights come from the stationary distribution of the transition probability matrix according to which neighbors are chosen at every time instant. 

\bibliographystyle{IEEEtran}
\bibliography{BayesRef}

\end{document}